# Realization and Topological Properties of Third-Order Exceptional Lines Embedded in Exceptional Surfaces

Weiyuan Tang[1,2], Kun Ding[3*], Guancong Ma[1*]

[1] Department of Physics, Hong Kong Baptist University, Kowloon Tong, Hong Kong, China

[2] Department of Physics, The University of Hong Kong, Pokfulam Road, Hong Kong, China

[3] Department of Physics, State Key Laboratory of Surface Physics, and Key Laboratory of Micro and Nano Photonic Structures (Ministry of Education), Fudan University,

Shanghai 200438, China

[*]e-mail: kunding@fudan.edu.cn; phgcma@hkbu.edu.hk

**Abstract**

As the counterpart of Hermitian nodal structures, the geometry formed by exceptional points (EPs), such as exceptional lines (ELs), entails intriguing spectral topology. We report the experimental realization of order-3 exceptional lines (EL3) that are entirely embedded in order-2 exceptional surfaces (ES2) in a three-dimensional periodic synthetic momentum space. The EL3 and the concomitant ES2, together with the topology of the underlying space, prohibit the evaluation of their topology in the eigenvalue manifold by prevailing topological characterization methods. We resolve this issue by defining a winding number that associates with the resultants of the Hamiltonian. This resultant winding number detects EL3 but ignores the ES2, allowing the diagnosis of the topological currents carried by the EL3, which enables the prediction of their evolution under perturbations. Our results exemplify unprecedented topology of higher-order exceptional geometries and may inspire new non-Hermitian topological applications.

*Introduction.*—Spectral degeneracies in band structure often possess intriguing topological properties. For example, in Hermitian three-dimensional (3D) systems, point degeneracies such as Dirac, Weyl, or triple points are monopoles of Berry flux [1,2], which determines the non-trivial and quantized cyclic evolution of wavefunctions. Degeneracies can form continuous geometries, e.g., nodal lines with intricate structures such as rings, links, and chains [3], which in some cases may even give rise to non-Abelian topological charges [4,5]. Nodal surfaces have also been shown to carry topological charges [6,7]. Recently, physicists found that introducing non-Hermiticity further enriches the diversity of band topology [8–11]. This is partly due to the fact that the non-Hermitian spectrum occupies the complex plane, such that the energies themselves can exhibit topological winding behaviors, leading to an additional layer of "spectral topology" underneath the wavefunction topology that is studied for Hermitian systems, giving rise to skin effects [12–16] and spectral knots [17,18]. Non-Hermitian degeneracies known as EPs possess topological properties characterizable by spectral winding numbers [8,10,11,19–21]. In most studies, EPs are formed by two coalescing states. These are called order-2 EPs, and akin to Hermitian degeneracies, they can also form nodal structures, such as rings [22–24], lines [25–27], links and chains [25,27,28], and surfaces [29,30]. Higher-order EP is formed when three or more states coalesce. They are intrinsically more stringent to realize, and their stable existence demands more degrees of freedom in the parameter space or a higher level of symmetries [31,32]. For these reasons, higher-order EP structures are so far beyond reach experimentally and are less well-understood theoretically.

Here, we report the experimental realization of EL3 entirely embedded on ES2. Both the EL3 and ES2 run continuously through the entire 3D parameter space, which is homeomorphic to a 3-torus by design. Such geometry presents an unexpected difficulty for topological characterization. The prevailing method extracts topological properties of nodal degeneracies by evaluating topological invariants either on a 2-sphere enclosing the entire nodal structure, with the topological charges of a Weyl point being an important case; or on a 1-sphere encircling the nodal structure, such as the characterization of topological nodal lines [3,7–9,33] and order-2 EP lines [21,24]. (Topological categorization using 0-sphere, i.e., two separate points, is also possible but not widely adopted. Hence it is not our focus here.) However, in our case, any enclosing sphere of a single EL3 would encounter ill-behaved spectral singularity on the ES2, thus defying the continuous requirement for spectral winding. This conundrum is resolved by exploring the resultants of the Hamiltonian matrix [34], which are auxiliary manifolds associated with but different from the eigenvalue manifolds. A "resultant vector field" can be chosen to vanish only at the EL3 and remain continuous at the ES2, through which a resultant winding number can be computed for diagnosing the topology of the EL3 while ignoring the influence of the ES2. The validity of our approach is verified by successfully predicting the local evolution of a touching point (TP) of two

EL3 under perturbation. Our study expands the understanding of non-Hermitian topology by unveiling novel topological scenarios exclusive to higher-order EP structures.

*Realization of symmetry-protected EL3.*—First, we present an experiment-feasible lattice system that realizes the said EL3. We begin with a codimension analysis of an order-$n$ EP (EP$n$), i.e., $n$-fold non-Hermitian degeneracy point. An isolated EP$n$ requires $2(n-1)$ degrees of freedom (DOFs) to have a solution. In the absence of any symmetry, a $d$-dimensional structure constituted by EP$n$ lives in the parameter space with minimal $2(n-1)+d$ dimensions. Hence both isolated EP3 and ES2 are stable in a four-dimensional (4D) parameter space [30,31]. Higher-dimensional systems have been studied in diverse scenarios using synthetic dimensions, such as acoustics [35–37], photonics [38–40], and electric circuits [41,42], they nevertheless tend to be more difficult to handle and realize. Fortunately, the dimensionality requirement can be reduced by enforcing additional symmetries [31]. In particular, when parity-time symmetry is respected, the characteristic polynomial of a Hamiltonian $H$, denoted $p(\omega) = \det(H - \omega \mathbf{I}) = 0$ where $\omega$ denotes the eigenvalues and $\mathbf{I}$ is an identity matrix, has entirely real discriminant $\mathcal{D} \in \mathbb{R}$, i.e., $\text{Im}\,\mathcal{D} = 0$ is always satisfied. Hence the DOF requirement is reduced from $2(n-1)$ to $n-1$. In other words, the codimension of EP2 is reduced by one and the codimension of EP3 by two [31,43]. Consequently, both ES2 and EL3 are accessible in a 3D PT-symmetric three-state system, serving as our starting point in designing an experiment-feasible lattice model.

We base our experimental system on coupled acoustic cavities [37,44]. Here, we further engineer the system such that its parameter space is mapped to a 3D lattice model. We begin with three air-filled cylindrical cavities stacked together [Fig. 1(a)]. Within each cavity, a thin plate is fixed in the radial direction, which reflects the circulating propagative waves and thus leads to the formation of azimuthal standing-wave modes. We use the second azimuthal mode, whose pressure $P$ and velocity $v$ profiles are shown in Figs. 1(b) and 1(c), respectively. The mode is harmonic in the azimuthal angle $\phi$, with a pressure node (velocity anti-node) along the diameter perpendicular to the plate. Such mode profiles inspire us to use the azimuthal angle to realize three $2\pi$-periodic synthetic coordinates, denoted $(\phi_1, \phi_2, \phi_3)$. Because the parameter space is clearly a homeomorphism of a 3-torus, we call it a 3D synthetic Brillouin zone (SBZ) henceforth. The azimuthal position of the nodal line is set as the "origin" of the three coordinates. Let $\phi_1$ tune the imaginary part of onsite resonant frequency in cavities A and C, which is the source of non-Hermiticity in our system. This is achieved by placing a piece of acoustic sponge to generate losses, which are linearly proportional to the local kinetic energy, given as $K \propto v^2(\phi_1) \propto \sin^2 \phi_1$. Let $\phi_2$ modulate the real part of the onsite resonant frequency of cavity B. A small metallic block is placed on the circumference for this purpose, and its azimuthal position is assigned as $\phi_2$. Its perturbation to the resonant frequency is linear to the local acoustic potential energy $U \propto P^2(\phi_2) \propto \cos^2 \phi_2$ [45]. Assign $\phi_3$ to control the coupling strength between cavities A and B, also B and C. Acoustic coupling strength is proportional to pressure intensity, i.e., $P^2(\phi_3) \propto \cos^2 \phi_3$. To satisfy PT symmetry, an equal amount of acoustic sponge to all three cavities as the biased loss, and then a specific amount of sponge in cavity A is relocated to C, such that an effective gain is created in A and the same amount of loss is added to C. We have characterized the detuning, loss, and coupling in our systems, and the results are shown in Fig. 1(d-f) as functions of the corresponding synthetic momenta. By delicately tuning the acoustic parameters of the whole ternary cavity system, we obtain that the tunable loss in the system follows $L(\phi_1) = -60.86 \times (0.50 \sin^2 \phi_1 - 1)$, the detuning of cavity B is described by $M(\phi_2) = -26.60 \cos^2 \phi_2$, and the coupling between neighboring cavities obeys $N(\phi_3) = -42.91 \times (1 - 0.90 \cos^2 \phi_3)$.

Our acoustic design is captured by a three-state Hamiltonian $H = (\omega_0 - i\gamma_0)\mathbf{I} + H_{3b}$, where $\omega_0$ is the resonance frequency of the second azimuthal cavity mode and $\gamma_0$ is the dissipation rate. The second term is

$$H_{3b}(\phi_1, \phi_2, \phi_3) = \begin{pmatrix} iL(\phi_1) & N(\phi_3) & 0 \\ N(\phi_3) & M(\phi_2) & N(\phi_3) \\ 0 & N(\phi_3) & -iL(\phi_1) \end{pmatrix}. \tag{1}$$

By using the trigonometric identity $\cos(2\phi) = 2\cos^2\phi - 1 = 1 - 2\sin^2\phi$, we obtain $L(\phi_1) = \gamma + 2\kappa_1 \cos(2\phi_1)$, $M(\phi_2) = \epsilon + 2\kappa_2 \cos(2\phi_2)$, and $N(\phi_3) = \beta + 2\kappa_3 \cos(2\phi_3)$. It then follows that model (1) maps to a periodic lattice shown in Fig. 2(a). Herein, the constant parameters are the onsite gain (loss) rate of site-a (c) $\gamma = -44.55$, the onsite offset to site-b $\epsilon = -18.90$, $\beta = -29.40$, $\kappa_1 = -7.42$, $\kappa_2 = -9.45$, and $\kappa_3 = 6.30$, all have the unit of rad/s and are obtained by benchmarking the experimental system. The non-Hermiticity in the model comes from the function $\pm iL(\phi_1)$, which manifests as the constant gain (loss) $\pm i\gamma$ and the asymmetric long-range hopping $\pm i2\kappa_1 \cos(2\phi_1)$.

The emergence of an EP$n$ can be identified by the conditions $\mathcal{R}[p^{(j)}, p^{(j+1)}] = 0$ with $0 \leq j < n - 1$, and $\mathcal{R}$ denotes the resultant, $p^{(j)}$ is the $j$th-order derivative of the characteristic polynomial with respect to $\omega$. As such, we identify both ES2 and EL3 in the SBZ, as shown in Fig.

2(b). The real-eigenfrequency Riemann surfaces on three distinct 2D slices are displayed in Fig. 2(c-e). Panels (d) and (e) show $\phi_1\phi_2$-planes sliced at $\phi_3 = \pi/2$ and $\phi_3 = \pi/3$, respectively. EL2 are observed as the consequence of the slicing planes cutting across the ES2, as highlighted by the dashed blue curves. The remaining state (shown in orange) touches the EL2 at particular isolated points and forms EP3 (purple hexagons and red stars). These EP3 only appear when $\phi_2 = \pm\pi/2$, or equivalently $M(\phi_2) = 0$. The conditions for EP3 to appear are $M(\phi_2) = 0$, $L(\phi_1) \pm \sqrt{2}N(\phi_3) = 0$, where the $\pm$ sign suggests two possible solutions (see Supplementary Materials, Section I for details). The EL3 are plotted in Fig. 2(b). They are also seen in Fig. 2(c), which plots the $\phi_1\phi_3$-plane at $\phi_2 = \pi/2$. Herein, the EL3 forms a linear crossing at $\phi_1 = 0, \pm\pi$, which we denote as touching points (TPs) and mark by the purple hexagons. The characteristics of the EL3 and the TPs will be discussed later.

The ES2 and EL3 are observed in our acoustic experiments. To this end, the acoustic pressure responses at each cavity are measured near $\omega_0$ at different synthetic momenta. The real and imaginary parts of the eigenfrequencies are then extracted from the acoustic responses using the Green's function [37,44,46] (see Supplementary Materials, Section II for details). We fix $\phi_2 = 0.5\pi$, then choose five different $\phi_3$ indicated by the horizontal dashed lines in Fig. 3(a), and for each $\phi_3$, the acoustic system is tuned to five different $\phi_1$. The real parts of the eigenfrequencies from the measured data are depicted in Fig. 3(b), which shows good agreement with the theoretical results (solid curves). Therein, the EP3 can be easily identified at positions where all three eigenfrequencies converge. These positions are then marked in Fig. 3(a) with the stars and fall on the computed EL3. We then observe the ES2 by performing similar experiments in different $\phi_1\phi_2$-planes at $\phi_3 = 0.5\pi$ and $\phi_3 = 0.33\pi$, which intersect with the ES2 and the EL3. The coalescence of two of the three states or all three states is clearly seen in Fig. 3(d, f), and the measured locations of the EP2s and EP3s also conform well with the theoretical results.

*Characterization of the EL3.*—The presence of both ES2 and EL3 gives rise to intriguing topological characteristics. The ES2 form close, continuous 2D surfaces that kiss at the TPs, which separate the eigenvalue manifold into disjoint regions. The EL3 are entirely embedded in the ES2 and also osculate at the TPs. Also, both the ES2 and EL3 run through the entire SBZ in the $\phi_1$-direction. Such peculiar geometries entail difficulties in their topological characterization. As mentioned before, the topology of a nodal structure is diagnosed by invariants computed on the $m$-spheres with $0 \leq m \leq d - 1$, which enclose the nodal structures. Examining the ES2 and EL3, it is clear that no 2-sphere can enclose them. Yet it remains possible to encircle both the ES2 and EL3 with a 1-sphere. We have computed the eigenvalue winding number, defined as $\mathcal{W} = \sum_{\mu \neq \nu} \left[ \frac{1}{2\pi i} \oint_{S^1} (d\vec{\phi} \cdot \nabla_\phi \arg(\omega_\mu - \omega_\nu)) \right]$ with $\mu, \nu$ indexing the states, which is a topological invariant for the spectral topology of the eigenvalue manifold. The result is $\mathcal{W} = 0$, which does not reveal the topological details carried by the ES2 or EL3 individually.

We thus must find an approach to characterize the topology of the EL3, which seemingly refuses any prevailing characterization methods. To its resolve, recall the discriminant of $p(\omega)$, defined as $\mathcal{D} = \prod_{\mu < \nu}(\omega_\mu - \omega_\nu)$. The equation $\mathcal{D} = 0$ indicates two or more identical eigenvalues and is commonly used as the condition for identifying an EP$n$. However, importantly, this condition does not distinguish $n$. The ensuing winding of $\mathcal{D}$, called the discriminant number widely used to characterize EP2, is equivalent to $\mathcal{W}$, which reflects the intersection multiplicity of $\text{Re}(\mathcal{D}) = 0$ and $\text{Im}(\mathcal{D}) = 0$ [20]. This enlightens that the quantity defined to model the EL3 shall vanish only at the EP3 to avoid the influence of the inevitably surrounded EP2s. The fact that $\mathcal{D} = -\mathcal{R}[p, p^{(1)}]$ rules out $\mathcal{R}[p, p^{(1)}]$, and the other two resultants, $\mathcal{R}[p, p^{(2)}]$ and $\mathcal{R}[p^{(1)}, p^{(2)}]$, shall be used because they fulfill the above requirement. Because generically $\mathcal{R}[p^{(k)}, p^{(j)}] \in \mathbb{C}$, the resultants themselves form a complex manifold that complements the eigenvalue manifolds, and there is a one-to-one correspondence between the zeros on the resultant manifold with EPs [34]. The winding of the resultants around these zeros directly associates with the winding of eigenvalues around the corresponding EPs. For example, the discriminant number is equivalent to the winding number for $\mathcal{R}[p, p^{(1)}]$. However, because the model (1) obeys PT symmetry, such that $\mathcal{R}[p^{(k)}, p^{(j)}] \in \mathbb{R}$, they do not exhibit any winding behavior individually. We circumvent this problem by further defining a complex function,

$$\Lambda(\vec{\phi}) \coloneqq \eta + i\zeta, \text{ with } \eta = \mathcal{R}[p^{(1)}, p^{(2)}], \zeta = \mathcal{R}[p, p^{(2)}]. \tag{2}$$

This way, $\Lambda$ vanishes only at EP3 and completely ignores EP2 (see Supplementary Materials, Section III for the proof).

In Fig. 4(b), we plot the $\Lambda$ as a vector field on the $\phi_1\phi_2$-plane at $\phi_3 = 0.7\pi$ [the green plane in Fig. 4(a)], which intersects with two EL3. $\Lambda$ is indeed vanishing at the EP3, but it does not generate any vortex. Consequently, the winding numbers of $\Lambda$, defined as

$$\mathcal{W}_\Lambda = -\frac{1}{2\pi} \oint_{S^1} \nabla_\phi (\arg \Lambda) \cdot d\vec{\phi}, \tag{3}$$

are zero for both EP3, suggesting that they both are unstable. To further reveal the local evolution of the EP3, we introduce two types of symmetry-preserving perturbations ($\delta_L$ and $\delta_M$): $L(\phi_1) = -60.86(1 - 0.50 \sin^2 \phi_1 + \delta_L)$ and $M(\phi_2) = -26.60\cos^2\phi_2 + \delta_M$. When perturbation is off, i.e., $\delta_M = \delta_L = 0$, the two surfaces defined by $\eta = 0$ and $\zeta = 0$ are respectively shown by orange and blue surfaces in Fig. 4(a). According to Bezout's theorem, the number of intersection points is four in $\phi_1\phi_2$-plane when considering both the complex domain and intersection multiplicity. However, in Fig. 4(b), only two intersections are found, which are the EP3s. This indicates a two-fold multiplicity for both intersections. Further changing $\phi_3$ to $\pi/2$, two EL3 merge and form the TP, which clearly has a multiplicity of four.

The two-fold multiplicity of the EL3 combined with their vanishing $\mathcal{W}_\Lambda$ together suggests that the EL3 in Fig. 4(a) can be made locally stable without breaking the symmetry. The speculation is easily verifiable by letting either $\delta_M$ or $\delta_L$ be non-zero. Figure 4(c) shows that when $\delta_M = -0.1$, the EL3 split into two pairs symmetric about the $\phi_2 = \pi/2$ plane, and they do so without dropping the order. Figure 4(d) plots the $\Lambda$-field and the solutions for $\eta = 0, \zeta = 0$ in the $\phi_1\phi_2$-plane at $\phi_3 = 0.7\pi$. Clearly, four EP3 are seen, indicating the removal of multiplicity. The EP3s can be separated into two pairs by the opposite vortices they carry, indicating $\mathcal{W}_\Lambda = \pm 1$. In other words, the EL3s are now topologically stable. Note that the TPs from the crossings of the two oriented order-3 ELs possess zero $\mathcal{W}_\Lambda$, and their multiplicity is reduced to two. We can use $\mathcal{W}_\Lambda$ to assign each EL3 with a "topological current," as indicated by the arrows in Fig. 4(c). Indeed, the currents cancel when the two pairs of EL3 merge. Herein, an alternative description is that EL3 carrying opposite topological currents can merge without annihilation, giving rise to higher-order EL that are "topologically neutral." Such a scenario, which resembles the accidental degeneracy appearing in Hermitian band structures, has not been reported before.

The topological currents are also informative in revealing the local evolution of the TPs. Such a configuration discloses two possible local evolutions in the natural projective plane ($\phi_1\phi_3$ plane). When the TP is open, the two linear-crossed EL3s can only separate without violating the orientation defined by the currents. The two possible cases are shown in Figs. 4(e) and 4(f).

*Discussion and conclusion.—* We have experimentally realized EL3 embedded in ES2 across the entire 3D SBZ. The discovery of such unconventional exceptional geometry reveals a multitude of intriguing aspects of higher-order EP previously unknown. The topological characterization of the EL3 demands escaping to an auxiliary resultant manifold, which remains well-behaved at the ES2 and only detects the EL3. Our findings show that higher-order EPs possess far richer topological properties that are absent for both EP2 and Hermitian degeneracies. The exploration of these properties may lead to new phenomena and applications relating to non-Hermitian energy transfer [47,48] or wave manipulations [49].

*Acknowledgment.*—This work is supported by the National Natural Science Foundation of China (11922416, 12174072), the Hong Kong Research Grants Council (RFS2223-2S01, 12302420, 12300419, 12301822), and the Natural Science Foundation of Shanghai (No. 21ZR1403700).

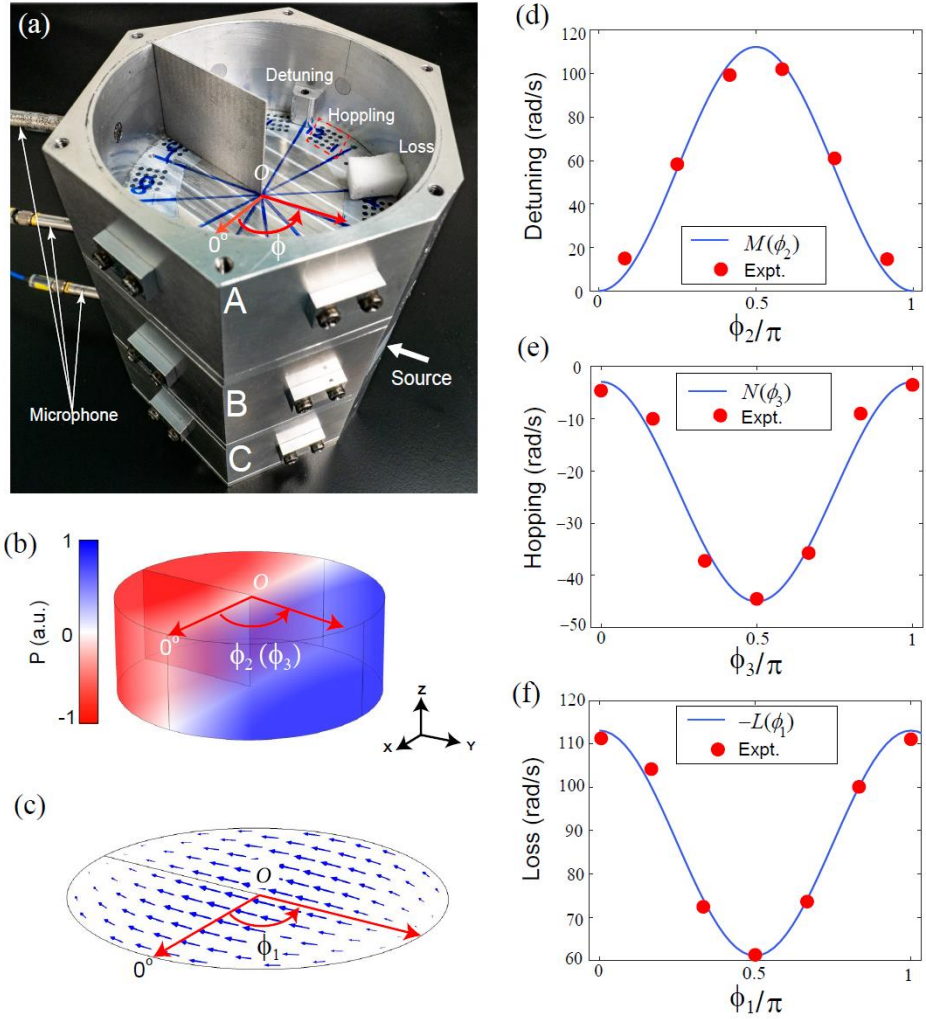

FIG.1 Experimental realization of the synthetic space. (a) A ternary acoustic cavity system realized the synthetic space. Each cavity has a height of 4 cm and a radius of 5 cm. (b-c) Acoustic pressure (b) and velocity field (c) of the second-order mode for a single cavity. (d-f) The onsite detuning, hopping strengths, and onsite loss as functions of $\phi_2$, $\phi_3$, and $\phi_1$, respectively. The blue curves are fitted from experimental data (red circles).

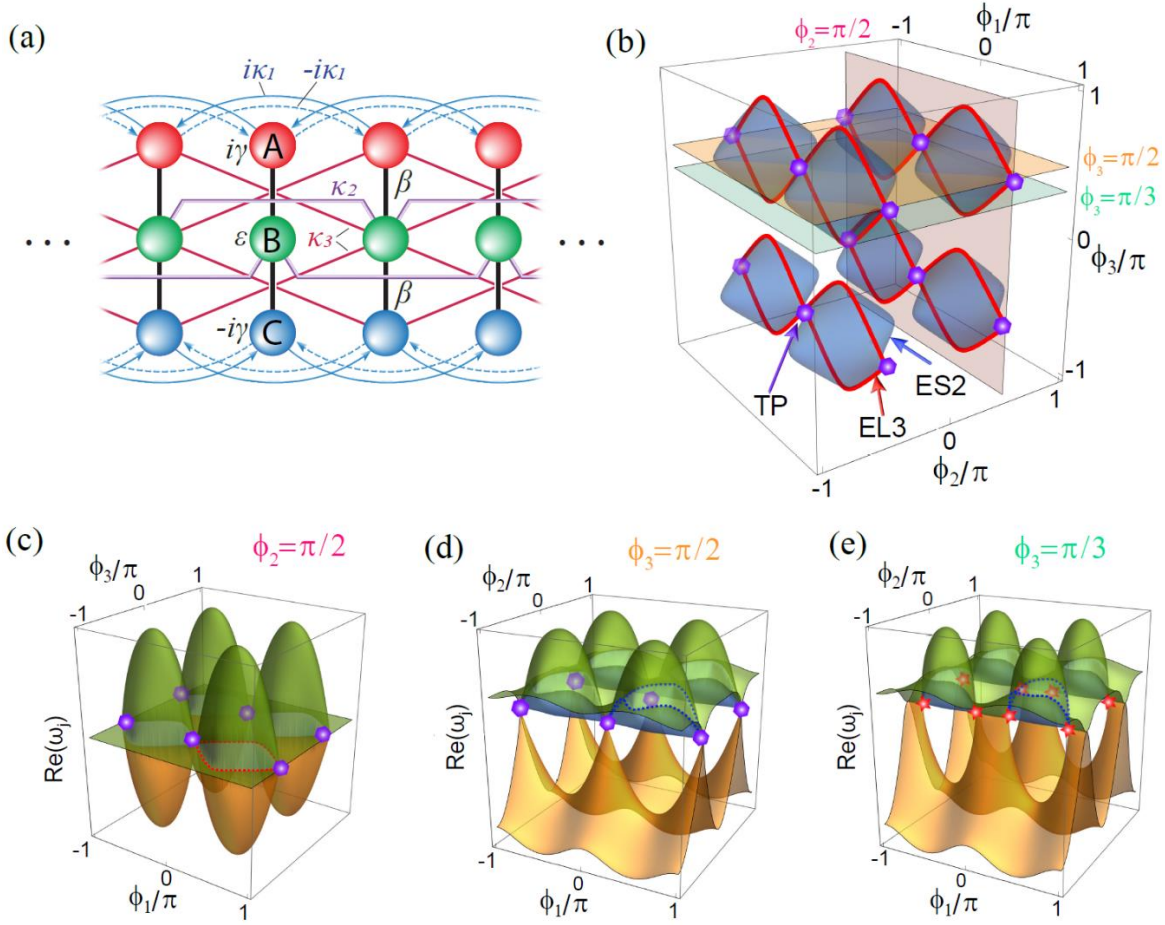

FIG. 2 (a) The lattice that maps to our model Eq. (1). (b) The EL3 (red curves) and ES2 (blue surfaces) in the SBZ. The purple hexagons denote the TPs of the EL3. (c-e) Real-eigenfrequency Riemann surfaces in the $\phi_2 = \pi/2$ plane (c), $\phi_3 = \pi/3$ plane (d), and $\phi_3 = \pi/2$ plane (e). The EP3s and EL3 are denoted by the red stars and dashed curves, respectively. The blue dashed curves show the EL2, which are cuts of the ES2.

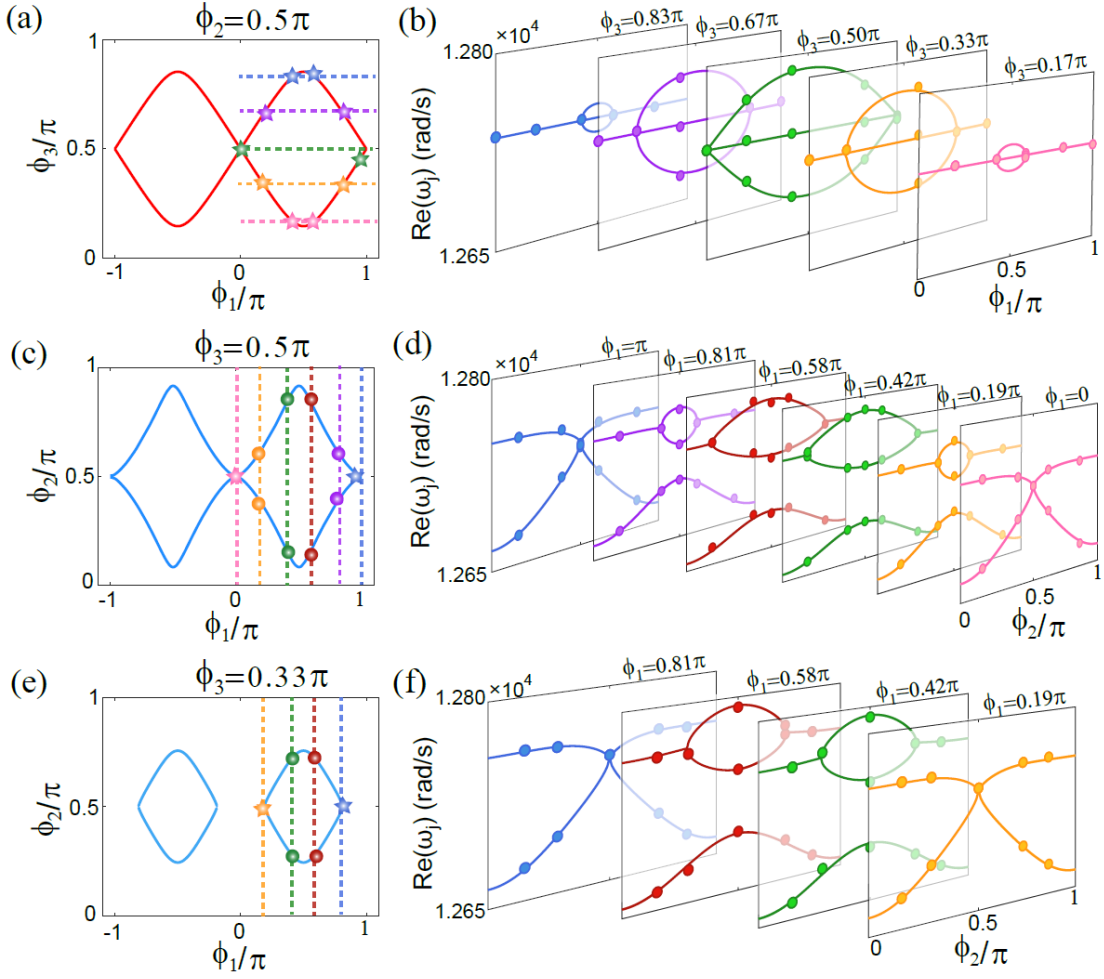

FIG. 3 Experimental results. (a) The EL3 in the $\phi_2 = 0.5\pi$ plane. (b) The measured real parts of eigenfrequencies along the dashed lines in (a). (c-f) Two slices of the ES2 at (c) $\phi_3 = 0.5\pi$ and (e) $\phi_3 = 0.33\pi$. The measured real parts of eigenfrequencies along the dashed lines in (c) and (e) are respectively shown in (d) and (f). The filled circles in (b), (d), and (f) represent the experimental measurement. The stars and circles in (a), (c), and (e) denote experimentally observed positions of EP3 and EP2.

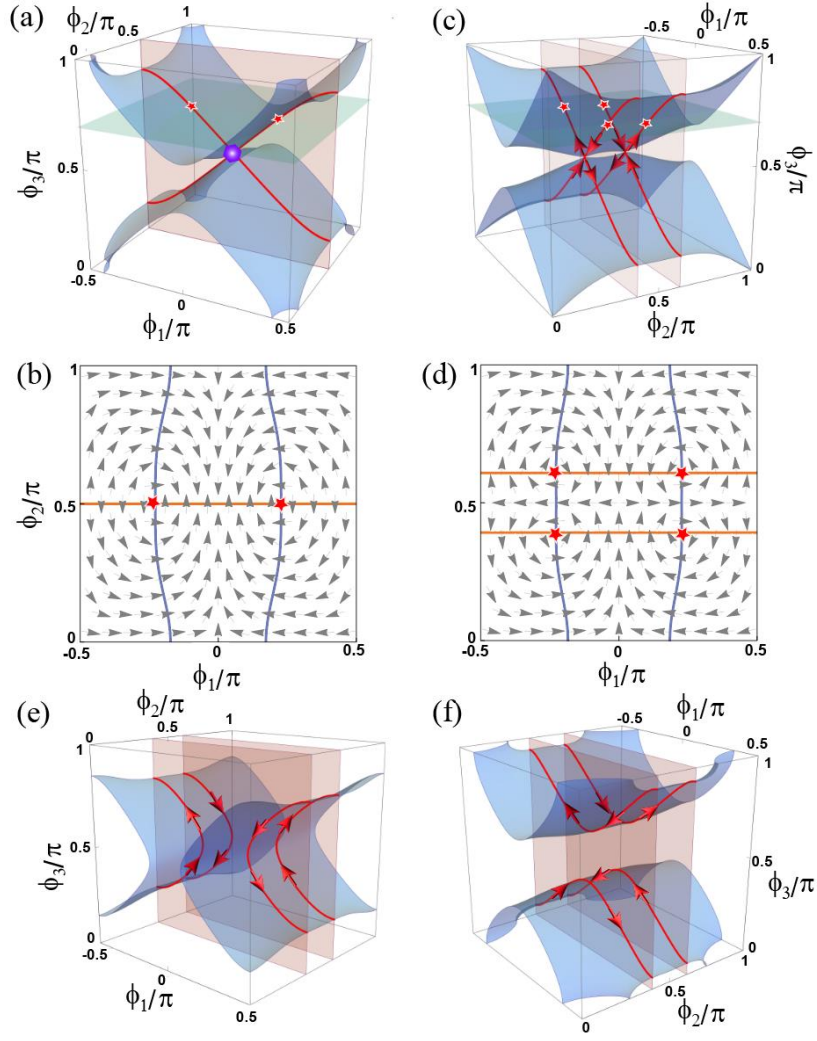

FIG. 4 Topological characterization of the EL3. (a) Topologically neutral EL3. The blue and red surfaces stand for $\text{Re}(\Lambda) = 0$ and $\text{Im}(\Lambda) = 0$. The purple dot is the TP. (b) The $\Lambda$-field (arrows) in of $\phi_3 = 0.7\pi$ plane [the green plane in (a)]. The EP3 (red stars) are found at the intersections of the blue line [$\text{Re}(\Lambda) = 0$] and the orange line [$\text{Im}(\Lambda) = 0$]. (c) With $\delta_L = 0$ and $\delta_M = -0.1$. The EL3 split into two pairs with opposite topological currents. (d) The $\Lambda$-field in the $\phi_3 = 0.7\pi$ plane. (e-f) The opening of the TPs under different perturbations: (e) $\delta_L = 0.1$, $\delta_M = 0$; (f) $\delta_L = -0.1$, $\delta_M = 0$.

# Supplemental Materials

# "Realization and Topological Properties of Third-Order Exceptional Lines Embedded in Exceptional Surfaces"

Weiyuan Tang[1,2], Kun Ding[3*], Guancong Ma[1*]

[1] Department of Physics, Hong Kong Baptist University, Kowloon Tong, Hong Kong, China

[2] Department of Physics, The University of Hong Kong, Pokfulam Road, Hong Kong, China

[3] Department of Physics, State Key Laboratory of Surface Physics, and Key Laboratory of Micro and Nano Photonic Structures (Ministry of Education), Fudan University,

Shanghai 200438, China

## I.     The locations of EP3s

The condition of EPs given by the vanishing of the EPs are found at the intersections of the surfaces defined by $\text{Re}(\mathcal{D}) = 0$ and $\text{Im}(\mathcal{D}) = 0$, where $\mathcal{D}$ is the discriminant of the characteristic polynomial $p(\omega) = \det(\omega \mathbf{I} - H_{3b})$. For our model,

$$p(\omega) = \omega^3 - M(\phi_2)\omega^2 + [L^2(\phi_1) - 2N^2(\phi_3)]\omega - L^2(\phi_1)M(\phi_2). \qquad (1)$$

For convenience, we denote $a_3 = 1$, $a_2 = -M(\phi_2)$, $a_1 = L^2(\phi_1) - 2N^2(\phi_3)$, and $a_0 = -L^2(\phi_1)M(\phi_2)$. Then, we have [1]

$$\mathcal{D} = (-1)^{N(N-1)/2} \mathcal{R}[p(\omega), p(\omega)^{(1)}] = (-1)^{N(N-1)/2} \det[\text{Syl}[p(\omega), p(\omega)^{(1)}]], \qquad (2)$$

wherein $N = 3$ for our system, $\text{Syl}[p(\omega), p(\omega)^{(1)}]$ is the Sylvester matrix

$$\text{Syl}[p, p^{(1)}] = \begin{pmatrix} a_3 & a_2 & a_1 & a_0 & 0 \\ 0 & a_3 & a_2 & a_1 & a_0 \\ 3a_3 & 2a_2 & a_1 & 0 & 0 \\ 0 & 3a_3 & 2a_2 & a_1 & 0 \\ 0 & 0 & 3a_3 & 2a_2 & a_1 \end{pmatrix}, \qquad (3)$$

wherein $a_i$ is the coefficients in $p(\omega)$ with $i = 0,1,2,3$. Explicitly,

$$\mathcal{D} = -L^6(\phi_1) + 8N^6(\phi_3) - 2L^4(\phi_1)M^2(\phi_2) - L^2(\phi_1)M^4(\phi_2) + 6L^4(\phi_1)N^2(\phi_3)$$
$$-10L^2(\phi_1)M^2(\phi_2)N^2(\phi_3) - 12L^2(\phi_1)N^4(\phi_3) + M^2(\phi_2)N^4(\phi_3). \qquad (4)$$

Notably, due to PT symmetry, $\mathcal{D} \in \mathbb{R}$, i.e., we have $\text{Im}(\mathcal{D}) = 0$ everywhere. Thereby, the codimension for EP2s can be reduced from two to one, while the codimension for EP3s decreases from four to two, which enables the EP2s and EP3s to simultaneously occur in a three-dimensional synthetic space. When $\phi_2 = \pm\pi/2$, which means $M(\phi_2)$ vanishes, the coalescence of three eigenstates gives rise to an EP3 [2]. Plug in $\phi_2 = \pm\frac{\pi}{2}$ to Eq. (1) yields

$$p(\omega) = \omega^3 + [L^2(\phi_1) - 2N^2(\phi_3)]\omega. \tag{5}$$

The discriminant $\mathcal{D}$ then vanishes at

$$L^2(\phi_1) - 2N^2(\phi_3) = 0. \tag{6}$$

Equation (6) can also be derived from the formulas of complex eigenvalues of the Hamiltonian. When $\phi_2 = \pm\pi/2$, the three eigenvalues are

$$\omega_{1,3} = (\omega_0 - i\gamma_0) \pm \sqrt{2N^2(\phi_3) - L^2(\phi_1)}, \quad \omega_2 = (\omega_0 - i\gamma_0). \tag{7}$$

Notably, three eigenvalues coalesce when the term under the square root in Eq. (7) vanishes, which recovers Eq. (6). The EL3s obtained from Eq. (6) are plotted in Fig. 2(a) in the main text. Herein, $\mathcal{D}$ contains the information of both EP2s and EP3s. To further distinguish EP3s from EP2s, we need to find the exclusive condition for EP3s.

The universal form of the characteristic polynomial $p(\omega) = \det(\omega I - H_{3b})$ is [3]

$$p(\omega) = \omega^3 - \text{tr}[H_{3b}]\omega^2 + \frac{(\text{tr}[H_{3b}])^2 - \text{tr}[H_{3b}^2]}{2}\omega - \det[H_{3b}], \tag{8}$$

and the corresponding discriminant is

$$\mathcal{D} = -\frac{1}{27}[4\xi^3 + \nu^2], \tag{9}$$

where $\xi$ and $\nu$ are both complex-valued constraints when symmetry is absent,

$$\xi_{3b} = \frac{1}{2}(\text{tr}[H_{3b}]^2 - 3\text{tr}[H_{3b}^2]), \tag{10}$$

$$\nu_{3b} = \frac{1}{2}(54\det[H_{3b}] - 5\text{tr}[H_{3b}]^3 + 9\text{tr}[H_{3b}]\text{tr}[H_{3b}^2]). \tag{11}$$

The subscript "3b" denotes the three-band case. When $N = 3$, Eq. (2) is further simplified as $\mathcal{D} = -\mathcal{R}[p, p^{(1)}]$. The other two resultants are also defined

$$\mathcal{R}[p, p^{(2)}] = \det[\text{Syl}(p, p^{(2)})] = \frac{1}{16}(54\det[H_{3b}] - 5\text{tr}[H_{3b}]^3 + 9\text{tr}[H]\text{tr}[H_{3b}^2]). \tag{12}$$

$$\mathcal{R}[p^{(1)}, p^{(2)}] = \det[\text{Syl}(p^{(1)}, p^{(2)})] = \frac{1}{24}(\text{tr}[H_{3b}]^2 - 3\text{tr}[H_{3b}^2]). \tag{13}$$

We can find

$$\xi_{3b} = \frac{1}{12}\mathcal{R}[p^{(1)}, p^{(2)}], \tag{14}$$

$$v_{3b} = \frac{1}{8}\mathcal{R}[p, p^{(2)}]. \tag{15}$$

Then relation among the three resultants is

$$\mathcal{R}[p, p^{(1)}] = \frac{1}{27}\left[\frac{1}{432}\mathcal{R}[p^{(1)}, p^{(2)}]^3 + \frac{1}{64}\mathcal{R}[p, p^{(2)}]^2\right]. \tag{16}$$

Remarkably, $\mathcal{R}[p, p^{(1)}]$ vanishes only when $\mathcal{R}[p^{(1)}, p^{(2)}] = 0$ and $\mathcal{R}[p, p^{(2)}] = 0$. Since the EP2s are determined by $\mathcal{R}[p, p^{(1)}] = 0$, both $\mathcal{R}[p^{(1)}, p^{(2)}] = 0$ and $\mathcal{R}[p, p^{(2)}] = 0$ fails to detect any EP2s. In a similar fashion, $\mathcal{R}[p^{(1)}, p^{(2)}]$ and $\mathcal{R}[p, p^{(2)}]$ can pinpoint EP3s while ignoring EP2s.

## II.  Retrieval of eigenfrequencies

The eigenfrequencies are obtained from the measured response spectra by utilizing the Green's function [3,4], given by

$$\overleftrightarrow{G}(\omega) = \sum_{j=1}^{3} \frac{|\psi_j^R\rangle\langle\psi_j^L|}{\omega - \omega_j}, \tag{17}$$

where $\omega_j$ denotes the eigenfrequency with $j$ labeling the eigenstates. The normalized biorthogonal right and left eigenvectors are respectively represented by $|\psi_j^R\rangle$ and $\langle\psi_j^L|$. Since the Hamiltonian in our work is reciprocal, we have $\langle\psi_j^L| = |\psi_j^R\rangle^T$. Besides, the pressure responses measured from three coupled acoustic cavities satisfy

$$P(\omega) = \langle m|\overleftrightarrow{G}(\omega)|s\rangle, \tag{18}$$

wherein $|s\rangle$ and $|m\rangle$ are 3×1 column vectors, which represent the source and microphone positions, respectively. In our experiment, the source pumps at cavity B, which is represented as $|s\rangle = (0\ \ 1\ \ 0)^T$. The pressure responses are measured at the central position of three cavities, thus $|m\rangle$ are $(1\ \ 0\ \ 0)^T$, $(0\ \ 1\ \ 0)^T$, and $(0\ \ 0\ \ 1)^T$ for cavities A, B, and C, respectively. Then the parameters of the Hamiltonian can be retrieved from the measured pressure response spectrum by applying a genetic algorithm. Some fitting results are selected and plotted in Fig. S1.

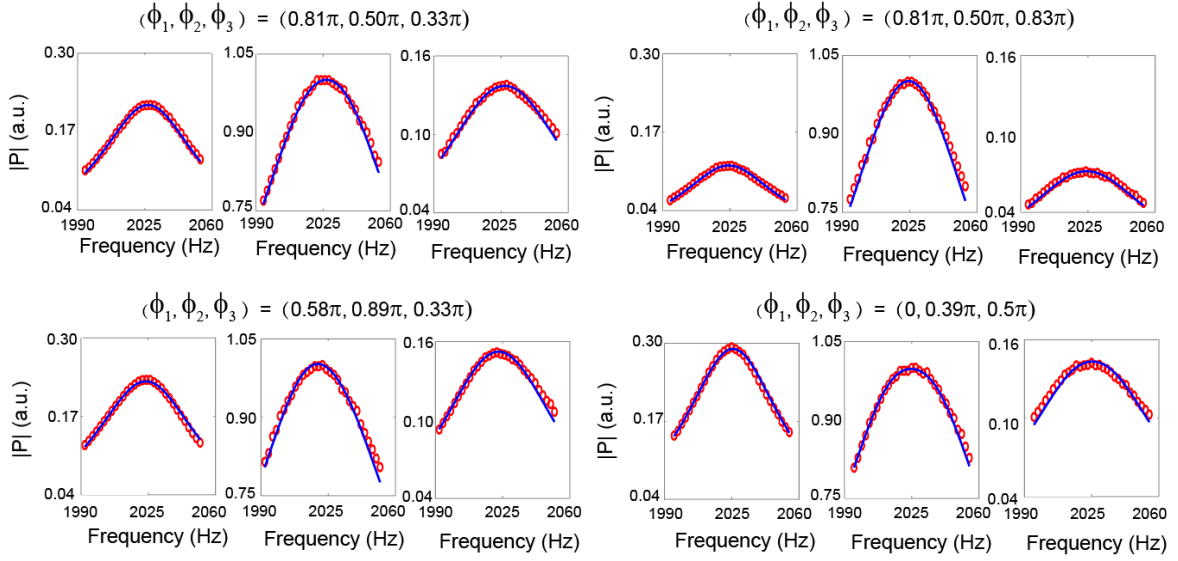

FIG. S1. Selected results of the measured and retrieval acoustic pressure response spectra. Red open circles are experimentally attained data. Blue curves are fitted by applying the Green's function method. Good agreement between the two indicates the validity of our fitting approach.

### III. The multiplicity of EP3s

The resultant field $\Lambda(\vec{\phi}) = \eta + i\zeta$ vanishes at EP3s, which corresponds to the intersections $\eta = 0$ and $\zeta = 0$, as shown in Fig. S2(a). Hence the multiplicity of the intersection is precisely the multiplicity of the EP3. For our Hamiltonian $H_{3b}$, $\eta$ and $\zeta$ are

$$\eta = -144 \begin{bmatrix} (\delta_M + 0.62 \cos^2 \phi_2)^2 + 6(-1 + 0.90 \cos^2 \phi_3)^2 \\ +6(1 + \delta_L - 0.50 \sin^2 \phi_1)^2 \end{bmatrix}, \quad (19)$$

$$\zeta = -128(\delta_M + 0.62 \cos^2 \phi_2) \begin{bmatrix} 9(1 + \delta_L - 0.50 \sin^2 \phi_1)^2 + \\ 9(-1 + 0.90 \cos^2 \phi_3)^2 + (\delta_M + 0.62 \cos^2 \phi_2)^2 \end{bmatrix}. \quad (20)$$

$\eta$ and $\zeta$ are real when $\delta_L$, $\delta_M \in \mathbb{R}$, which can also be understood from the symmetry perspective. Since the second bracket in Eq. (20) is always positive, $\zeta = 0$ can be deduced from

$$\delta_M + 0.62 \cos^2 \phi_2 = 0. \quad (21)$$

This implies that when $\frac{\delta_M}{0.62} > 0$, no EP3 exists. When $\frac{\delta_M}{0.62} < 0$, the EP3s appear around $\phi_2 = \pm \pi/2$. Substituting Eq. (21) into Eq. (19) gives the location of EL3s

$$0.50 \sin^2 \phi_1 - 0.90 \cos^2 \phi_3 = \delta_L, \quad (22)$$

$$0.50 \sin^2 \phi_1 + 0.90 \cos^2 \phi_3 = 2 + \delta_L. \quad (23)$$

Concerning our current experimental setup, only the EL3s predicted by Eq. (22) exist. Therefore, $\eta = 0$ in Eq. (19) can be rewritten as

$$[0.50 \sin^2\phi_1 - 0.90 \cos^2\phi_3 - \delta_L][2 - 0.50 \sin^2\phi_1 - 0.90 \cos^2\phi_3 + \delta_L] = 0. \quad (24)$$

Then $\eta$ and $\zeta$ vanish when

$$\delta_M + 0.62 \cos^2\phi_2 = 0, \quad (25)$$

$$0.50 \sin^2\phi_1 - 0.90 \cos^2\phi_3 - \delta_L = 0. \quad (26)$$

According to Bezout's theorem, the number of intersection points with multiplicity is four in both $\phi_1\phi_2$ and $\phi_2\phi_3$ planes. As shown in Fig. S2(a-c), the number of EP3s is indeed four in both $\phi_1\phi_2$ and $\phi_2\phi_3$ planes.

Note that when $\delta_M = \delta_L = 0$, we only find a single TP at $\left(0, \frac{\pi}{2}, \frac{\pi}{2}\right)$. The multiplicity of this TP can be revealed by further expanding $\phi_1$, $\phi_2$, $\phi_3$ near $\left(0, \frac{\pi}{2}, \frac{\pi}{2}\right)$ as $\sin^2\phi_1 \cong q_1^2, \cos^2\phi_2 \cong q_2^2, \cos^2\phi_3 \cong q_3^2$, then Eqs. (25) and (26) are simplified as

$$0.62\, q_2^2 = 0, \quad (27)$$

$$0.50\, q_1^2 - 0.90\, q_3^2 = 0. \quad (28)$$

Equations (27) and (28) tell the multiplicity of the TP at $\left(0, \frac{\pi}{2}, \frac{\pi}{2}\right)$ is also four in both $q_2q_3$ and $q_1q_2$ planes. When $\delta_M \neq 0$ and $0.90 \cos^2\phi_3 + \delta_L \neq 0$, there are four EP3s falling on the $\phi_1\phi_2$ plane or $\phi_2\phi_3$ plane that cuts the EL3. Take the $\phi_1\phi_2$ plane at $\phi_3 = 0.7/\pi$ shown in Fig. S2(a) for example. To simplify the discussion, we further expand $\phi_1$, $\phi_2$ around the EP3 encircled by the purple closed loop, which arises as the intersection $P$ of the curves of $\zeta = 0$ and $\eta = 0$. Then the conditions for the vanishing of $\zeta$ and $\eta$ are rewritten as

$$F_\zeta = q_2, \quad F_\eta = q_1. \quad (29)$$

Obviously, when $q_1 = q_2 = 0$, $F_\zeta$ and $F_\eta$ intersect at $P_{\eta,\zeta}$ with multiplicity $u_{P_{\eta,\zeta}}(F_\eta, F_\zeta) = 1$. Notably, EP3 can also be identified as the intersections of either two out of the three resultants (recall that $\eta = \mathcal{R}[p^{(1)}, p^{(2)}], \zeta = \mathcal{R}[p, p^{(2)}]$, and we further denote $\chi \coloneqq \mathcal{R}[p, p^{(1)}]$ for convenience). The conditions given by $\eta = 0$ and $\zeta = 0$ are already discussed as the resultant field $\Lambda = \eta + i\zeta$. But we still have two more possibilities: $\chi = 0$ and $\zeta = 0$, or $\chi = 0$ and $\eta = 0$, as shown in Fig. S2(b, c). We investigate the multiplicities of these two intersections, denoted $P_{\chi,\zeta}$ and $P_{\chi,\eta}$. From Eq. (2) and Eq. (9), we can deduce $\chi = \frac{1}{27}\left(\frac{1}{432}\eta^3 + \frac{1}{64}\zeta^2\right)$. Near the EP3, the condition of $\chi = 0$ is

$$F_\chi = \frac{1}{27}\left(\frac{1}{432}q_1^3 + \frac{1}{64}q_2^2\right). \tag{30}$$

Remarkably, the tangent of $F_\chi$, denoted as $m(F_\chi)$, is $q_2$, which is of multiplicity 2. $F_\zeta$ also has one tangent of $q_2$ with the multiplicity being 1, but $F_\eta = q_1$ does not have any linear term in $q_2$, leading to $m(F_\eta) = 0$. Then we have

$$u_{P_{\chi,\zeta}}(F_\chi, F_\zeta) = m(F_\chi) + m(F_\zeta) = 3, \tag{31a}$$

$$u_{P_{\chi,\eta}}(F_\chi, F_\eta) = m(F_\chi) + m(F_\eta) = 2, \tag{31b}$$

$$u_{P_{\eta,\zeta}}(F_\eta, F_\zeta) = m(F_\eta) + m(F_\zeta) = 1. \tag{31c}$$

Apparently, only $u_{P_{\eta,\zeta}}(F_\eta, F_\zeta)$ exactly characterize the multiplicity of EP3. This is because $F_\chi$ is also the condition of EP2 and has singularities at the positions of EP3s. Hence one cannot use $F_\chi$ to characterize the multiplicity of the EP3. To further demonstrate, we define two different resultant fields $\Lambda' = \chi + i\zeta$, and $\Lambda'' = \chi + i\zeta$. They are plotted in Fig. S2(b) and (c). The corresponding resultant windings are shown in Fig. S2(d-f). When the intersection $P_{\eta,\zeta}$ is encircled, we obtain $\mathcal{W}_\Lambda = 1$. However, encircling intersection $P_{\chi,\zeta}$ and $P_{\chi,\eta}$ gives $\mathcal{W}_{\Lambda'} = 1$ and $\mathcal{W}_{\Lambda''} = 0$, respectively, resulting from the singularity existing on the curve $F_\chi$, which coincides with their multiplicities.

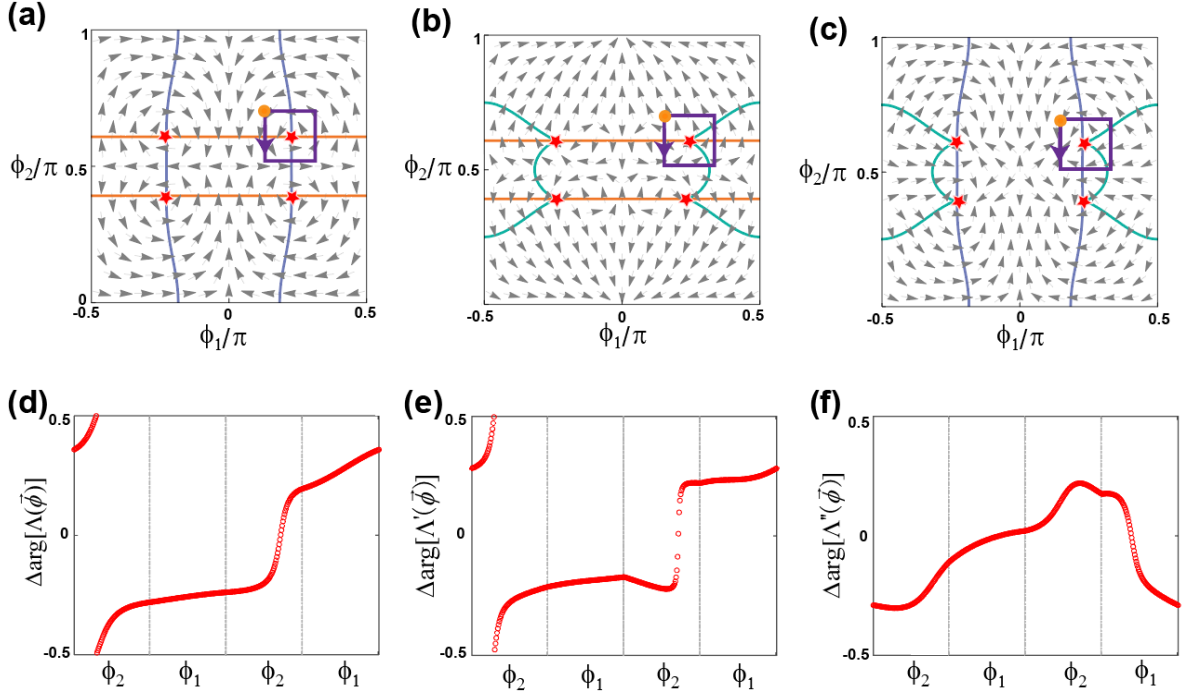

Fig. S2. The resultant fields (arrows) in the $\phi_1\phi_2$-plane at $\phi_3 = 0.7\pi$, as denoted by the green plane labeled in Fig. 4(a). Here, the fields are $\Lambda = \eta + i\zeta$ for (a), $\Lambda' = \chi + i\zeta$ for (b), and $\Lambda'' = \chi + i\eta$ for (c). The evolutions of the argument when encircling the EP3s along the purple closed loops in (a-c) are plotted in (d-f). In (a-c), $F_\eta$, $F_\zeta$ and $F_\chi$ are represented by blue, red and green curves, respectively. The red stars denote EP3s. The orange points indicate the start and finish points, with the purple arrows denoting the direction of encircling.